\documentclass[twocolumn,showpacs,nofootinbib,10pt]{revtex4-1}

\usepackage[colorlinks]{hyperref}
\usepackage{graphicx} 
\usepackage{color,float,amssymb,amsmath,upgreek}
\definecolor{green1}{RGB}{0,128,0} 
\hypersetup{hidelinks,backref=true,pagebackref=true,hyperindex=true,colorlinks=true,breaklinks=true,urlcolor= blue}
\hypersetup{%
  colorlinks = true,
  linkcolor  = blue,
  citecolor = green1,
}
\usepackage{bookmark,textgreek}
\usepackage{hyperref,color,xcolor}
\hypersetup{hidelinks,hyperindex=true,colorlinks=true,breaklinks=true,urlcolor= blue}
\hypersetup{%
  colorlinks = true,
  linkcolor  = blue
}\usepackage{upgreek}

\newcommand{\beq}{\begin{eqnarray}}
\newcommand{\benu}{\begin{enumerate}}
\newcommand{\enu}{\end{enumerate}}
\newcommand{\eeq}{\end{eqnarray}}

\begin{document}

\title{The emergence of flagpole and flag-dipole fermions in fluid/gravity correspondence}

\author{P. Meert}
\email{pedro.meert@ufabc.edu.br} 
\affiliation{Federal University of ABC, Center of Natural and Human Sciences, Santo Andr\'e, Brazil}
\author{R. da Rocha}
\email{roldao.rocha@ufabc.edu.br}
\affiliation{Federal University of ABC, Center of Mathematics, Computing and Cognition, Santo Andr\'e, Brazil}


\begin{abstract}
The emergence of flagpole and flag-dipole singular spinor fields is explored, in the context of fermionic sectors of fluid/gravity correspondence, arising from the duality between the gravitino, in supergravity, and the phonino, in supersymmetric hydrodynamics. Generalized black branes, whose particular case consists of the AdS--Schwarzschild black brane, are regarded.  The correspondence between  hydrodynamic transport coefficients, and the universal absorption cross sections of the generalized black branes, is extended to fermionic sectors, including supersound diffusion constants. 
A free parameter, in the generalized black brane solution, is shown to control the flipping between regular and singular fermionic solutions of the equations of motion for the gravitino.
 \end{abstract}
\maketitle

\section{Introduction}
Classical spinor fields were classified studying all the possibilities to evaluate their respective bilinear covariants that either satisfy the Fierz identities or their generalizations. This feature  has introduced the Lounesto's spinor field classification into six classes of spinor fields, assuming the U(1) gauge symmetry of quantum electrodynamics \cite{Lounesto:2001zz}. A second quantized
 version of such a classification was introduced  in Ref.  \cite{Bonora:2017oyb}, where quantum spinors and their correlators provided a setup  for a second quantized classification. 
Going further, encompassing ${\rm SU}(2)\times {\rm U}(1)$ gauge symmetry, a new classification,  embracing spinor field  multiplets that represent non-Abelian gauge fields, was lately  introduced in Ref.   \cite{Fabbri:2017lvu}.  Recently, new classes of fermionic fields on  7-manifolds were derived \cite{Bonora:2015ppa,Bonora:2014dfa}, regarding, in particular, the AdS$_5\times S^5$ and AdS$_4\times S^7$   compactifications \cite{Martinho:2018oxl}, also including new fermionic solutions of M-theory compactifications with one
supersymmetry \cite{Lopes:2018cvu}. These new classes emulate singular spinor fields on higher dimensions and more general signatures. 
 Hence, it is natural to further explore the role of the spinor fields classifications in the fluid/gravity correspondence setup. 

 The low-energy/low-momentum limit of the AdS/CFT correspondence is also known as fluid/gravity correspondence. In this regime, the field theory side is taken to be an effective theory, hence, hydrodynamics \cite{Landau:ThV06}. On the other hand, the compactification of higher dimensions leads the gravitational theory to conventional General Relativity (GR), although this gives some freedom to play with extensions of GR and investigation of its dual theories. This fact has lead to successful predictions of transport coefficients in strongly coupled field theories, being the quark-gluon plasma \cite{Finazzo:2013efa}  the most famous example, but not the only one, also appearing in other setups, like the graphene \cite{Lucas:2015sya}, superconductors \cite{Hartnoll:2008kx}, and Fermi liquids \cite{Cubrovic:2009ye}. One intriguing feature of this duality is the so called KSS result \cite{Policastro:2001yc,KSS}, named after Kovtun, Son and Starinets, which states that the shear viscosity to entropy density ratio is universal, in the sense that its numerical value is the same for almost all known physical systems. One exception involve a highly complex framework \cite{Brigante:2007nu}. 

To lead the fluid/gravity correspondence -- essentially based on bosonic fields -- further, one aims to include fermionic modes into the description. To accomplish so, one refers to supersymmetry in the bulk and analyzes its effect in the boundary, describing supersymmetric hydrodynamics \cite{Kovtun:2003vj}. This setup indeed leads to predictions \cite{SSLee:2009,HLiu:2011,MCubrovic:2009,TFaulkner:2011,Hartnoll:2009,Iqbal:2009} and the quest which concerns us in this work is related to the problem of whether a quantity similar to the shear viscosity to entropy ratio, associated to fermionic sectors, exists. In Ref. \cite{Policastro:2008cx} the sound diffusion constant was first calculated in a supersymmetric holographic background and indicated that this quantity is the obvious candidate for the task, which was investigated and asserted later by \cite{Erdmenger:2013thg}.

The sound diffusion constant is related to the super current, which turns out to be the super partner of the energy-momentum tensor. When one considers a field theory with $T>0$, supersymmetry is spontaneously broken and the emergence of a collective fermionic excitation called phonino arises in very general circumstances \cite{Lebedev:1989rz}. The sound diffusion is associated with the damping of this mode, i. e., imaginary part of the dispersion relation. It was computed analytically for $\mathcal{N}=4$ supersymmetry. In the holographic setting the EOM for the gravitino in AdS$_5$ background were solved to first order in the frequency and momentum using the retarded green function of the dual supersymmetric current, from where the dispersion relation was read off.

Flagpoles and flag-dipoles are types of the so called singular spinor fields in Lounesto's U(1) gauge  classification. Flagpoles encompass neutral Majorana, and Elko, spinor fields, as well as  
charged spinor fields satisfying specific 
Dirac equations \cite{daRocha:2005ti}.
Flag-dipoles are very rare in the literature, being their first appearance in Ref. \cite{esk}. 
The emergence of flagpole and flag-dipole singular spinor fields in the context of fermionic sectors of fluid/gravity correspondence is here scrutinized, exploring the duality between the graviton, in a supergravity bulk setup, and the phonino, in the boundary supersymmetric hydrodynamics. These spinor fields emerge when generalized black branes are considered, whose particular case in the AdS--Schwarzschild black brane for a very particular choice of parameter.
This parameter appearing in the generalized black branes shall be shown to drive the flipping that takes regular into  singular spinors fields, as solutions of the equations of motion for the gravitino.

This paper is organized as follows: Sect. II is devoted to a brief review of the U(1) spinor field classification from bilinear covariants. The Fierz identities, and their generalizations, are discussed as well as the role of singular and regular spinor fields. In Sect. III, the relation between hydrodynamic transport coefficients and the universal absorption cross sections in the corresponding gravity dual is provided, and  then extended to the fermionic sectors. The Kubo formul\ae\, for the gravitino transport coefficients is used. Sect. IV is dedicated to explore the aforementioned results for generalized black branes and derive the supersound diffusion constants. The bulk action for the gravitino is employed in the generalized black brane background, showing how flagpole and flag-dipole fermionic fields emerge as solutions of the derived equations of motion. The free parameter, in the generalized black brane, is then analyzed, also providing 
the flipping between regular and singular fermionic fields.

\section{General Bilinear Covariants and Spinor Field Classes}
A classical spinor field $\uppsi$ is an object of the irreducible representation space of the Spin group. In $1+3$ dimensions, the isomorphism 
Spin($1,3)\simeq$ SL(2,$\mathbb{C}$) means that a classical spinor field then carries the 
 representations of the SL(2,$\mathbb{C}$) Lorentz group. The bilinear covariants components, defined at each point $x$ on a 4D spacetime, with cotangent basis $\{e^\mu\}$  read
\begin{subequations}
\begin{eqnarray}
\Upsigma(x) &=& \overline{\uppsi}(x)\uppsi(x)\,,\label{sigma}\\
J_{\mu }(x) &=&\overline{\uppsi}(x)\upgamma _{\mu }\uppsi(x)\,,\label{J}\\
S_{\mu \nu }(x) &=&i\overline{\uppsi}(x)\upgamma
_{\mu
\nu }\uppsi(x),\label{S}\\
 K_{\mu}(x) &=&i\overline{\uppsi}(x)\upgamma_{0123}\upgamma _{\mu }\uppsi(x)\,,\label{K}\\\Upomega(x)&=&-\overline{\uppsi}(x)\upgamma_{0123}\uppsi(x)\,,  \label{fierz}
\end{eqnarray}\end{subequations}
where $\overline\uppsi=\uppsi^\dagger\upgamma_0$ is the spinor conjugate, $\upgamma
_{\mu
\nu }=\frac{i}{2}[\upgamma_\mu, \upgamma_\nu]$, 
$\upgamma_5=i\upgamma_{0123}=i\upgamma_0\upgamma_1\upgamma_2\upgamma_3$ is the volume element, and $\upgamma_{\mu }\upgamma _{\nu
}+\upgamma _{\nu }\upgamma_{\mu }=2\eta_{\mu \nu }\mathbb{I}_{4\times 4}$, being the $\upgamma_\mu$  the Dirac gamma matrices and the $\eta_{\mu\nu}$ are the Minkowski metric components.  The form fields ${\bf J}(x) = J_\mu(x)\,e^\mu$,   ${\bf K}(x) = K_\mu(x)\,e^\mu$, and ${\bf S}(x) = \tfrac12 S_{\mu\nu}(x)\,e^\mu\wedge e^\nu$ are defined, where $\alpha\wedge\beta$ denotes the exterior product of form fields $\alpha, \beta$. Exclusively in the Dirac electron theory, the 1-form ${\bf J}$ represents a U(1) conserved current density. More precisely, in natural units, the time component $J_0$ is well known to regard the charge density, and the spatial components  $J_i$ typifies the electric current density. 
The spatial components $S_{jk}$ represent the magnetic dipole moment density, whereas the mixed components, $S_{i0}$, denote the electric dipole moment density. The $K_\mu$ denote the chiral current density, that is solely conserved  in the massless case. The  scalar $\Upsigma$, responsible for the mass term in a fermionic  Lagrangian, and the pseudoscalar $\Upomega$, that is capable to probe CP symmetries, can be composed as $\Upsigma^2+\Upomega^2$ to be interpreted as a probability density. These  interpretations hold for the more common and usual cases where, for instance, the spinor describes the electron in the Dirac theory. Further cases can borrow similar interpretations for at least some of the bilinear covariants if the Dirac equation is satisfied for the given spinor field \cite{Villalobos:2015xca}. 

The U(1) classification of spinor fields is described by the following  classes \cite{Lounesto:2001zz},\begin{subequations}\beq
\!\!\!\!\!\!\!\!\!\!\!\!1)\;\;&&\,\; \Upsigma(x)\neq0\neq\Upomega(x),\;\;\;\mathbf{S}(x)\neq 0\neq \mathbf{K}(x),\label{tipo1}\\
\!\!\!\!\!\!\!\!\!\!\!\!2)\;\;&&\,\; \Upsigma(x)\neq0,\;\;\;
\Upomega(x) = 0,\;\;\;\mathbf{S}(x)\neq 0\neq\mathbf{K}(x),\label{tipo2}\\
\!\!\!\!\!\!\!\!\!\!\!\!3)\;\;&&\,\;\Upsigma(x)= 0, \;\;\;\Upomega(x) \neq0,\;\;\;\mathbf{S}(x)\neq 0\neq\mathbf{K}(x),\label{tipo3}\\
\!\!\!\!\!\!\!\!\!\!\!\!4)\;\;&&\,\;\Upsigma(x)=0=\Upomega(x),\;\;\;\mathbf{S}(x)\neq 0\neq\mathbf{K}(x),
\label{tipo4}\\
\!\!\!\!\!\!\!\!\!\!\!\!5)\;\;&&\,\;\Upsigma(x)=0=\Upomega(x),\;\;\;\mathbf{S}(x)\neq0,\;\;\;
\mathbf{K}(x)=0,
\label{tipo5}\\
\!\!\!\!\!\!\!\!\!\!\!\!6)\;\;&&\,\;\Upsigma(x)=0=\Upomega(x),\;\;\; \mathbf{S}(x)=0,
\;\;\; \mathbf{K}(x)\neq0.\label{tipo6}\eeq\end{subequations}
\noindent  

When both the scalar and the pseudoscalar vanish, a spinor field is called singular, otherwise its said to be regular. The objects in Eqs. \eqref{J} and \eqref{K}, being 1-form fields,  are named poles. Since spinor fields in the class 4, \eqref{tipo4}, have non vanishing {\bf K} and {\bf J}, 
spinor fields in this class are called flag-dipoles, because  ${\bf S}\neq 0$ is a 2-form field, identified by a flag, according to Penrose.
Besides, spinor fields in class 5, \eqref{tipo5}, have a vanishing pole, ${\bf K} = 0$, a non null pole, ${\bf J}\neq0$, and a non null flag, ${\bf S}\neq 0$, being  flagpoles. Spinor fields in class 6, \eqref{tipo6}, present two poles, ${\bf J}\neq0$ and ${\bf K}\neq0$, and a null flag, ${\bf S}=0$, corresponding therefore to a flag-dipole.  Flag-dipole spinor fields were shown to be a legitimate solution of the Dirac field equation in a torsional setup \cite{esk,Fabbri:2010pk,Vignolo:2011qt}, whereas Elko \cite{Ahluwalia:2009rh} and Majorana uncharged spinor fields represent type-5 spinors \cite{daRocha:2005ti}, although 
a recent example of a charged flagpole spinor has been shown to be a solution of the Dirac equation. Additional spinor fields classes, upwards of the Lounesto's classification, 
complete all the formal possibilities, including ghost fields  \cite{Villalobos:2015xca}. 
 The standard, textbook, Dirac spinor field is an element of the set of regular spinors in the class 1, \eqref{tipo1}. Besides, chiral spinor fields were shown to 
 correspond to be elements of the class 6, \eqref{tipo6}, of (dipole) spinor fields. Chiral spinor fields governed by the Weyl equation are Weyl spinor fields. However, the class 6 of dipole spinor fields further  allocates  
mass dimension one spinors, whose dynamics, of course, is not ruled by the Weyl equation, as well as flagpole spinor fields in the class 5, that are not neutral and satisfy the Dirac equation. The spinor field class 5, still, is also composed by mass dimension one spinor fields \cite{Ahluwalia:2009rh,Fabbri:2010ws,daRocha:2005ti}. 
The Lounesto's spinor field classification was also explored in the lattice approach to quantum gravity  \cite{Ablamowicz:2014rpa}. Flipping between regular and singular spinor fields was scrutinized in Ref. \cite{Cavalcanti:2014uta}, for very special cases.

The classification of spinor fields, according 
to the bilinear covariants, must not be restricted 
to the U(1) gauge symmetry of quantum electrodynamics. In fact, a more general classification, based on the ${\rm SU}(2)\times {\rm U}(1)$ gauge symmetry, embraces   multiplets and provide new fermionic possibilities in the electroweak setup  \cite{Fabbri:2017lvu}.

Regular spinor fields satisfy the Fierz identities, 
\begin{subequations} 
\begin{eqnarray}\label{fifi}
\!\!\!\!\!\!\!\!\!\!\!\Upomega S_{\mu\nu}+\Upsigma \epsilon_{\mu\nu}^{\quad\rho\sigma}S_{\rho\sigma}&=&K_\mu{J}_\nu,\\
\Upomega^{2}+\Upsigma^{2} &=&{J}^\mu J_\mu,\\{J}^\mu{K}_\mu 
&=&0={K}_\mu K^\mu +J_\mu J^\mu,\label{3c}
\end{eqnarray}
\end{subequations}
what does not hold, in general, for singular spinor fields. Notwithstanding, a multivector field,  constructed upon the bilinear covariants, 
\begin{equation}
\mathfrak{Z}(x)=\Sigma(x)+\mathbf{J}(x)+i\mathbf{S}(x)-\upgamma_{5}\mathbf{K}(x)+\upgamma_{5}\Omega(x),
\label{boom1}
\end{equation}
 is a {Fierz aggregate} if the listed bilinear covariants obey the Fierz identities (\ref{fifi} -- \ref{3c}). Besides,  Fierz aggregates,  that are  self-adjoint under the Dirac conjugation, are named    {boomerangs} \cite{Lounesto:2001zz}.
For singular spinor fields, the Fierz identities (\ref{fifi}) are promoted to the generalized ones,  \begin{subequations}\beq
\label{nilp}\mathfrak{Z}^{2}(x)  &=&4\Sigma(x) \mathfrak{Z}(x),\\
\mathfrak{Z}(x)\upgamma_{\mu}\mathfrak{Z}(x)&=&4J_{\mu}(x)\mathfrak{Z}(x),\\
\mathfrak{Z}(x)i\upgamma_{\mu\nu}\mathfrak{Z}(x)&=&4S_{\mu\nu}(x)\mathfrak{Z}(x),\\
\mathfrak{Z}(x)\upgamma_{5}\upgamma_{\mu}\mathfrak{Z}(x)  &=&4K_{\mu}(x)\mathfrak{Z}(x),\\
 \mathfrak{Z}(x)\upgamma_{5}\mathfrak{Z}(x)&=&-4i\Omega(x) \mathfrak{Z}(x),\label{nilp1}
\eeq\end{subequations}
that are satisfied for all the spinor fields in the Lounesto's classification. Moreover, spinor fields may be reconstructed from 
bilinear covariants, resulting in a classification  of spinor fields that is mutual to the Lounesto's one \cite{Cavalcanti:2014wia,fabbri}. In fact, given a spinor field $\upxi$
satisfying $\overline\upxi\uppsi\neq0$, then $
\uppsi=\frac{1}{4a}e^{-i\upalpha}\mathfrak{Z}\upxi$,  where $4a^2={\overline\upxi\mathfrak{Z}\upxi}$ and
$e^{-i\upalpha}=\frac{1}{a}\overline\upxi\psi$ \cite{Lounesto:2001zz}.

In the next section,   the relation between hydrodynamic transport coefficients and the universal absorption cross sections in the corresponding gravity dual is provided, and  then extended to the fermionic sectors. A generalized black brane background shall be introduced, 
providing the first steps for the emergence of flagpole and flag-dipole singular spinor fields. The  hydrodynamic transport coefficients, in the 
fermionic sectors, shall be briefly reviewed. The supersound diffusion constant,  for the generalized black brane, shall be  also studied, leading to the AdS--Schwarzschild  results \cite{Policastro:2008cx,Kontoudi:2012mu}, in a very particular limit.

\section{Black hole absorption cross sections and fermionic sectors}

Hydrodynamics plays the role of an effective account of quantum field theories (QFTs) in the  long
wavelength regime~\cite{Bu:2014ena},  regulated by a  the local fluid variables that are near the equilibrium. 
Transport coefficients, encompassing viscosities and conductivities, drive perturbations propagation and  can be experimentally measured.
One of the most remarkable predictions of AdS/CFT and fluid/gravity correspondence
is the shear viscosity-to-entropy density ratio, which is universal for a large class
of isotropic, strongly coupled, plasmas~\cite{Bu:2014ena}.
The fact that the shear viscosity-to-entropy density ratio is universal occupies a featured role in gauge theories that are dual to
certain gravitational backgrounds~\cite{Policastro:2001yc,KSS}. The universality demonstrations of the KSS result,  
$ \frac{\eta}{s}=\frac{1}{4\pi}$,  
illustrate how the shear viscosity, $\eta$,  of the hydrodynamic limit of the QFT, with energy momentum tensor $T_{\mu\nu}$, is related to the low-energy absorption cross section $\sigma(\omega=0)$ of a transverse bulk graviton $h_{12}$ by a black brane \cite{Buchel:2003tz,Kovtun:2004de,Son:2007vk}. Comparing the Kubo formula, 
\begin{equation}
   \eta = \lim_{\omega\rightarrow 0} \frac{1}{2\omega}\int  \left\langle \left[T_{12}(x),T_{12}(0)\right]\right\rangle\,e^{i\omega t}\,d^4 x
\end{equation}
to the low-energy absorption cross section,  $\sigma(\omega)=-\frac{2\kappa^2}{\omega}  \Im  G^R(\omega)$~\cite{Klebanov:1997kc,Gubser:1997yh}, where $\kappa^2 = 8\pi G$ denotes the gravitational coupling constant, yields
\begin{equation}
   \sigma(\omega)=\frac{\kappa^2}{\omega}\int   \left\langle \left[T_{12}(x),T_{12}(0)\right]\right\rangle\,e^{i\omega t}\,d^4 x.\label{scs}
\end{equation}
The entropy, $S=\frac{A}{4G}$, of a black brane is only dependent on its area at the horizon. Denoting by $s$ and $a$ the respective entropic and areal densities, one may employ the Klein--Gordon equation of a massless scalar, as an equation of motion for the low-energy absorption cross section associated with $h_{12}$, yielding the horizon area density $\sigma(0)=a$ \cite{Erdmenger:2013thg}. 

Black branes in a 5D bulk have near-horizon geometry  
\begin{equation}\label{3bbb}
	ds^2=-N(r)dt^2 +A(r) dr^2+ \frac{r^2}{\ell^2} d\vec{x}_3^2,
\end{equation} 
where ${\ell}$ denotes the AdS$_5$ radius, which shall be considered unity herein, for the sake of conciseness. 
 At the strong coupling regime $g_s N_c \gg 1$, the branes considerably  
curve the background bulk, sourcing the  geometry of the generalized black brane \cite{Casadio:2016zhu},
\begin{subequations}
\beq\label{cfm1}
N(r) &=& r^2\left(1-\frac{r_+^4}{r^4}\right)
,\\
A^{-1}(r) &=& r^2\left[\left(1-\frac{r_+^4}{r^4}\right)\frac{\left(1-\frac{5r_+^4}{2\,r^4}\right)}{1-(4\beta+1)\frac{r_+^4}{2r^4}}\right],\label{cfm2}
\eeq
\end{subequations}
where $r_+$ denotes the black brane horizon. 
These generalized 
black branes can be equivalently obtained in two ways. The first one consists of deforming,  under the ADM formalism, the AdS--Schwarzschild black brane \cite{Casadio:2016zhu}. The second manner to derive them is as an analytical black brane solution of quadratic Ricci gravity 
with Lee--Wick terms. 
Clearly $\lim_{\beta\to 1} A(r) = N^{-1}(r)$, corresponding to the well known AdS--Schwarzschild black brane. 

The cross-section $ \sigma(0)=a$ was  previously proved  for 5D black brane metrics of type \cite{Das:1996we} 
\begin{equation}
 ds^2=-N(r)dt^2 + B(r)\left(dr^2 + r^2 d\Omega_3^2\right)\,.\label{cros1}
\end{equation}
In order to use the solutions (\ref{cfm1}, \ref{cfm2}), 
one needs to transform the metric \eqref{cros1}
into the standard one
\begin{equation}
 ds^2=-N(r)dt^2   + A(r)dr^2+ r^2 d\Omega_3^2,
\end{equation}
by introducing the variable $\mathring{r} = rB^{1/2}(r)$, yielding the expression
\begin{eqnarray}
A(\mathring{r}) = \left(\frac{B(r)}{B(r)+\frac{r}{2}B'(r)}\right)^2.
\end{eqnarray}
Therefore, one obtains 
\begin{widetext}
\begin{eqnarray}\label{mtcal}
B(r) &=&\frac{\left(1 \sqrt{\frac{2 r^8-5 r^4 r_+^4}{\left(r^4-r_+^4\right) \left(r^4-(4 \beta +1)
   r_+^4\right)}}- \sqrt{2} \sqrt{r^2-\frac{r_+^4}{r^2}}\right) \exp \left({  \frac{2r^2}{r_+^2}
     F\left(\left.i \sinh ^{-1}\left(r
   \sqrt{-\frac{1}{r_+^2}}\right)\right|-1\right)}
   \right)}{r^3 \sqrt{r^2-\frac{r_+^4}{2r^2}(\beta+1)} \sqrt{\frac{2 r^8-5 r^4
   r_+^4+1}{\left(r^4-r_+^4\right) \left(r^4-(4 \beta +1) r_+^4\right)}}},
\end{eqnarray}
\end{widetext}
where $F(\;\cdot\;|\;\cdot\;)$ denotes the elliptic  integral of the first kind. 
Ref.~\cite{Das:1996we} derived an analogue result for the low-energy absorption cross section of a massless,  minimally coupled, fermion by the black brane \eqref{cros1},
\begin{equation}
  \sigma_{\frac12}(0)=2\, (B(r_+))^{-3/2}\,a\,,\label{crossf}
\end{equation}
which is solely dependent on the generalized black brane event horizon \cite{Erdmenger:2013thg}. 

Transport coefficients in the usual fluid/gravity correspondence can be emulated by the fermionic sector of the theory. In fact,  the cross section can be  computed by the standard QFT rate of decaying particles,
$
   \sigma_{\frac12}=({2\omega})^{-1}\int  |\mathcal{M}|^2\,d\Uppi\,,
$ 
where the measure $d\Uppi$ accounts the momentum space of final state particles. One then considers a spinor field, describing an AdS$_{5}$ bulk fermion $\Uppsi$, whose kinetic part in the action reads $
 \int  \overline{\Uppsi}\upgamma^A{D}_A\Uppsi\,\sqrt{-g}\, d^{5}x,$ 
up to a constant, where $A=0,\ldots,4$, $\upgamma^A$ is a set of gamma matrices and the $D_A$ stand for the covariant derivative.  The notation $x^4=r$ shall be used to denote 
the AdS radial coordinate. 
Denoting by $\Uppsi_b$ the fermionic fields at the boundary, one considers their coupling to a spinorial  boundary operator, $S$, that accounts the transverse component of the supersymmetric current \cite{Antoniadis:1985az,Henningson:1998cd,Iqbal:2009fd}, by $
\int \left(\overline{S}P^-\Uppsi_b + \overline{\Uppsi}_bP^+S\right)\,d^4x $ \cite{Erdmenger:2013thg}. The chiral projector $P^\pm =\frac{1}{2}\left(1\pm\upgamma^{4}\right)$ is employed, where  $\upgamma^4$ denotes the gamma matrix corresponding to the radial AdS$_5$ coordinate.  Taking a fermion at rest in the boundary  implies that 
\begin{align}
   \sigma_{\frac12}(\omega)&=
   \frac{\kappa^2}{{\rm tr}_0}\text{ Tr}\left(-\upgamma^0 \upzeta\right), 
\end{align}
where $\upzeta =  \Im\int \left\langle \,P^-P^+ S(x) \overline{S}(0)  \,\right\rangle\,e^{i \omega t}\,d^4x$ and, hereon, we use the notation ${\rm tr}_0\equiv\text{ Tr}\left(-\upgamma^0 \upgamma^0\right)$.

The cross section can be, therefore, associated with the Kubo formul\ae \ for  coefficients of transport in the boundary  CFT \cite{Erdmenger:2013thg}. 
The supersymmetry current  $S^i$ is associated with the supercharge density, $\uprho = S^0$, by the so called constitutive relation
\cite{Erdmenger:2013thg}
\begin{equation}
   S^j = -\frac{P}{\upepsilon} \upgamma^0\upgamma^j \uprho +\left( \frac{{\rm D}_\upsigma}{2} [\upgamma^{j},\upgamma^{i}] - {\rm D}_{\rm s}\delta^{ij}\right)\nabla_i \uprho\,,\label{eq1230}
\end{equation}
where $\upepsilon$ and $P$ are the energy density and pressure of the fluid, respectively. Ref. \cite{Hoyos:2012dh} interprets $\uprho$ as a sound-like excitation, the phonino. The quantities  ${\rm D}_{\rm s}$ and ${\rm D}_\upsigma$ play the role of transport coefficients that govern the phonino, that has a speed dissipation given by $v_{\rm s}=\frac{P}{\upepsilon}$~\cite{Lebedev:1989rz,Leigh:1995jw}.  
Eq. (\ref{eq1230}) can be rewritten, taking into account to the spin-representations of O(3), 
\begin{equation}
\!\!\!\!   S^i \!=\!-\frac{P}{\upepsilon} \upgamma^0 \upgamma^i \uprho  \!-\!\left[ D_{\frac12} \upgamma^i\upgamma_j\!+\!D_{\frac32} \left(\delta^i_j\!-\!\frac{1}{3} \upgamma^i \upgamma_j\right)\right]\nabla^j \uprho\,,
\end{equation}
where $D_{\frac12} = \frac{1}{3} \left({\rm D}_{\rm s} -  {\rm D}_\upsigma\right)$ and $D_{\frac32} = {\rm D}_{\rm s} + \frac{1}{2} {\rm D}_\upsigma$, respectively, denote transport coefficients associated with the spin-$\frac12$ and the spin-$\frac32$  components of $\nabla^j \uprho$ \cite{Erdmenger:2013thg}. It emulates the splitting of the stress-energy tensor  into the shear,  $\eta$, and the bulk, $\zeta$, viscosities \cite{Policastro:2002tn,Herzog:2003ke}. In the conformal setup, it reads 
\begin{equation}
   {\rm D}_{\rm s} = \frac{2}{3} D_{\frac32} \quad\text{and}\quad D_{\frac12} =0\,.\label{constantesdifusao}
\end{equation}
Using the Kubo formul\ae\, yields \cite{Kontoudi:2012mu,Erdmenger:2013thg}
\begin{align}
    D_{\frac32}&=\frac{1}{\upepsilon\,{\rm tr}_0}\lim_{\omega,k\rightarrow 0}\text{ Tr}\left(-\upgamma^0 \upzeta_1\right)\,. \label{Kubo}
\end{align} One has $\upzeta_1=\Im\int \langle \,P^+ \d{S}^i(x) \overline{\d {S}}^i(0)P^- \,\rangle\,e^{i \omega t}\,d^4x$. The supersymmetric (transverse)  current  \begin{equation}
    \d {\rm S}^i\equiv\left(\delta^i_j - \frac{1}{3}\upgamma^i\upgamma_j\right)S^j,
\end{equation}
and the equation of conservation, $\partial_\mu S^\mu=0$, auxiliates to derive the correlator $\left\langle \uprho \overline{\uprho}\right\rangle$ as \cite{Erdmenger:2013thg}
\begin{equation}
\Im\,\left(\frac{k_i}{k^\mu k_\mu}\,\left\langle \d {\rm S}^i \overline{\uprho}\right\rangle\right)=-\frac23D_{\frac32}\,\Re\, \left\langle \uprho \overline{\uprho}\right\rangle\,.
\end{equation}
Therefore the gravitino can be coupled to the boundary supersymmetric current. The AdS/CFT correspondence 
 associates the gravitino absorption cross section with the dual operator Green's function \cite{Gubser:1997yh}. Similar to the way that one considers transverse metric perturbations $h_{12}$ for the shear viscosity-to-entropy ratio, one takes into account the gravitino modes that have spin-$\frac32$, namely, $\upupsilon_i=P_{ij}\Uppsi_j$, where the $P_{ij}$ operator   
is responsible to project onto the $\upgamma^i$ component a quantity transversal to $\vec{k}$. The transverse gravitino components then satisfy equations of motion associated with fermions. One may, hence,   associate the absorption cross section  to the Kubo formul\ae\,~\eqref{Kubo}, \begin{equation}
 D_{\frac32} = \frac{2}{ {\upepsilon}\,\kappa^2}\,\sigma_{\frac12}(0)\,,\label{kfsi}
\end{equation}
characterizing the fermionic version of the viscosity $
 \eta = \frac{1}{2\kappa^2}\,\sigma(0)$.

The cross section \eqref{crossf} 
can be further tuned, by considering a  regular fermion, \cite{Das:1996we,Erdmenger:2013thg} satisfying the Dirac equation $
\left(\upgamma^\mu\nabla_\mu  - m\right)\Uppsi =0\,.$ 
It is worth to mention that there are examples of 
regular and singular spinor fields, allocated in at least five of the six Lounesto's classes, that satisfy the Dirac equation.
Ref. \cite{Erdmenger:2013thg} studied the Dirac equation   
in the AdS--Schwarzschild background geometry. Taking into account the  generalized black brane geometry \eqref{cros1}, the Dirac equation yields 
\begin{equation}
 \left[k\upgamma^4\left(d_r\!+\!\frac{3}{2r}\right)\!+\!\frac{k}{r}\upgamma^j\text{\d{$\nabla$}}_j-\!m\sqrt{N}-\!i\omega \upgamma^0\right] \upxi\!=\!0,
\end{equation}
 where $d_r\equiv\frac{d}{dr}$, $k(r)=\sqrt{N(r)/B(r)}$, and one suitably scales 
 the fermionic field $\Uppsi$ by $\upxi =  \sqrt[4]{NB^{3}}\,\Uppsi$, 
 for easily solving the subsequent equations. Using an eigenspinor basis  of the operators $\{\upgamma^4,\upgamma^0\}$, namely, $\upgamma^{4}\lambda^\pm_n=\pm \lambda^\pm_n$ and $\upgamma^{0}\lambda^\pm_n=\mp \lambda^\mp_n$, the  expansion 
$
  \upxi =\sum_{n=0}^\infty F_n^\pm(r) \lambda_n^\pm, 
$
can be then employed \cite{Erdmenger:2013thg},  yielding  
  \begin{align}
  k\left(d_r + \frac{f_{n}^\pm}{r}\right)F_n^\mp \pm m \sqrt{N} F_n^\mp &= i \omega F_n^\pm\,,  \label{mde}
  \end{align}
denoting $f_{n}^+=n+3$, $f_{n}^-=-n$. 
Their $n=0$ fermionic mode  satisfies
\begin{equation}
\!\!\!\!\! \left[k\left(d_r+\frac{4}{r}+m\sqrt{B}\right)k\left(d_r - m\sqrt{B}\right) + \omega^2\right] F_0 =0\,.
\end{equation} 
Defining {\it x}-coordinates implicitly by $\frac{d}{dx} = k(r) \rho(r)r^3 \frac{d}{dr}$ \cite{Erdmenger:2013thg}, such that $d_r \rho=2m\rho\sqrt{B}$, implies that $\lim_{r\rightarrow \infty}\rho=  1$. Therefore, the $n=0$ fermionic mode can be rescaled as  $\mathfrak{F}_0 = e^{-m\int \sqrt{B}\,dr}F_0$, yielding 
\begin{equation}
 \left(\partial_x^2+(\omega \rho r^{3})^2 \right)\mathfrak{F}_0=0\,.
\end{equation}
Hence, the absorption cross section (per areal density) for a  massive spin-$\frac12$ fermion reads \begin{equation}
 \frac{\sigma_{\frac12}(0)}{a} = (B(r_+))^{-3/2} \exp\left(2m\int^{r_+}_\infty \sqrt{B}\,dr\right)\,.
\end{equation}
This result is led to the standard one derived in the context of the AdS--Schwarzschild black brane, in Refs. \cite{Das:1996we,Erdmenger:2013thg}, implemented in the limit  $\beta\to1$, in Eqs. (\ref{cfm1}, \ref{cfm2}, \ref{mtcal}). 
The supersound diffusion constant ${\rm D}_{\rm s}$, for the generalized black branes~\eqref{3bbb}, Eqs. \eqref{constantesdifusao} and~\eqref{kfsi}, then yields \cite{Policastro:2008cx,Kontoudi:2012mu}
\begin{equation}
 2 \, \pi T {\rm D}_{\rm s} =\frac{4\sqrt{2}}{9}\,, \label{sgen}
\end{equation}
where $T$ is the black brane temperature.

\section{Supersound diffusion constant from the transverse gravitino}\label{transverse}
Hereon the supersound diffusion constant
shall be studied with respect to the generalized black brane (\ref{cfm1}, \ref{cfm2}), also emulating the results for the AdS--Schwarzschild black brane in Ref. \cite{Erdmenger:2013thg} and the ones in Ref. \cite{Policastro:2008cx} for very particular limits. 
The bulk action for the gravitino,  
\begin{equation}
 S=\int \overline{\Uppsi}_\mu \left(\Gamma^{\mu\nu\rho}D_\nu-m\Gamma^{\mu\rho}\right)\Uppsi_\rho\,\sqrt{-g}\, d^{5}x,\label{bulk}
\end{equation} 
is employed, for $\mu,\nu,\rho=0,\ldots,4$,
The covariant derivative acts on spinors as $D_\mu = \partial_\mu + \frac{1}{4}\omega_\mu^{bc}\upgamma_{bc}$, where the $\omega_\mu^{bc}$ denotes the spin-connection, as usual. Employing the usual gauge condition $\Gamma^\mu \Uppsi_\mu=0$,  the Rarita--Schwinger equation reads  $
 \left(\upgamma^\mu {D}_\mu+m\mathbb{I}_{4\times 4}\right) \Uppsi_\mu =0\,.$
Supposing a boundary plane wave dependence $e^{-i \omega t + i k x}$ in the gravitino wave function, the projetor $P_{ij}$  is again used, to select the  gravitino components, $\upupsilon_i =\Uppsi_i - \frac{1}{2}\upgamma_i \upgamma^j \Uppsi_j$ ( $i,j\neq 1$). Hence,  the equations of motion, in the background~(\ref{3bbb}, \ref{cfm1}, \ref{cfm2}),  read 
\begin{equation}
\!\! \!\!\frac{\upupsilon^\prime}{\upupsilon} + \frac{\upgamma^5}{\sqrt{A}}\left(\frac{i k}{r} \upgamma^1\!-\!\frac{i \omega}{\sqrt{N}}  \upgamma^0\!-\! m\right)   \!+\!\frac{N^\prime}{4 \sqrt{NA}} \!+\! \frac{3}{2r}  \!=\!0\,.\label{eom_eta}
\end{equation}
Now, one writes \cite{Erdmenger:2013thg}
\begin{align}
\upupsilon = \upupsilon^{\upalpha+} \upalpha^++ \upupsilon^{\upalpha-} \upalpha^- +\upupsilon^{\vartheta+} \vartheta^++\upupsilon^{\vartheta-} \vartheta^-\,,\label{esplit}
\end{align}
where the basis of chiral ($\upalpha^+, \vartheta^+)$ and anti-chiral ($\upalpha^-, \vartheta^-)$ eigenspinors of the $i\upgamma^{1}\upgamma^2$ operator, respectively with $+1$ and $-1$ eigenvalues, concomitantly satisfying 
\begin{subequations}
\beq
&\upgamma^0 \upalpha^\pm=\pm \upalpha^\mp\,,\quad\upgamma^0 \vartheta^\pm = \mp \vartheta^\mp\,,\label{espin}\\&\upgamma^1 \upalpha^\pm = \pm i \, \vartheta^\mp\,,\quad\upgamma^1 \vartheta^\pm = \pm i\, \upalpha^\mp\,.
\eeq
\end{subequations}
 The correlator arises out of Eq.~\eqref{eom_eta} \cite{Son:2002sd,Policastro:2002se}, looking at the singular component  of Eq.~\eqref{eom_eta} near the event horizon, where the solutions $\upupsilon^{\upalpha\pm}$ are demanded to be of type \cite{Erdmenger:2013thg} 
\begin{equation}
  \upupsilon^{\upalpha\pm}\propto \left(r-r_+\right)^{-\frac{1}{4}\left(1- i\omega/\pi T\right)}\upupsilon^{\upalpha}_{0}\,,\label{solucoes}
\end{equation}

The supersound diffusion constant can be obtained from the Kubo formul\ae\, \eqref{Kubo}, used in the regime $\omega,k\to0$
 into~\eqref{eom_eta}, in the black brane background brane (\ref{cfm1}, \ref{cfm2}), yielding 
\begin{equation}
 \upupsilon^\pm =c^\pm \sqrt[8]{AN} \sqrt[3]{r^{-2}} \sqrt[4]{\sqrt{r}+r^2\sqrt{1-\frac{r^4}{r_+^4}}}^{\;\pm3}\,.\label{sol}
\end{equation} The phonino dispersion relation 
$
  \omega = v_{\rm s} k - i {\rm D}_{\rm s} k^2
$  can be then obtained from the pole of the longitudinal supersymmetry current correlator. The Rarita-Schwinger  equations read 
\begin{subequations}
\label{eom2}
\begin{align}
	&\!\!\!\!\!\!\!\upgamma^5 \Uppsi_0^\prime \!+\!\left(\!\frac{N^\prime}{4 \sqrt{A}} \upgamma^5\!-\! \frac{i \omega}{\sqrt{A}} \upgamma^0   \!+\! \frac{i k}{r} \upgamma^1 \!+\! \frac{3\sqrt{N}}{2 r} \upgamma^5 \!+\!{m\mathbb{I}_{4\times 4}}\right)\!\frac{\Uppsi_0}{\sqrt{N}}\nonumber\\&\quad- \frac{N^\prime}{2 \sqrt{AN}} \upgamma^0 \Uppsi_4 =0 \,,\\	
	&\!\!\!\!\!\!\!\upgamma^5 \Uppsi_4^\prime \!+\!\left[\left(\frac{N^\prime}{4 \sqrt{AN}}+\!\frac{5}{2r}\right) \upgamma^5 \!+\! \frac{ik}{r\sqrt{N}} \upgamma^1\! \!-\! \frac{i \omega}{\sqrt{AN}} \upgamma^0\!\right.\nonumber\\&\left.\quad+\frac{m\mathbb{I}_{4\times 4}}{\sqrt{N}} \right] \Uppsi_4  -\left(\frac{N^\prime}{2 \sqrt{AN}} \upgamma^0+ \frac{1}{r}\upgamma^0\right) \Uppsi_0  =0\,,\\
	&\!\!\!\!\!\!\!\upgamma^5 \Uppsi_j^\prime +\left(\frac{N^\prime}{4 \sqrt{AN}} \upgamma^5- \frac{i \omega}{\sqrt{AN}} \upgamma^0+ \frac{i k {}}{r\sqrt{N}} \upgamma^1 \right.\nonumber\\&\left.\qquad\qquad+ \frac{3}{2r} \upgamma^5+ \frac{m\mathbb{I}_{4\times 4}}{\sqrt{N}}\right)\Uppsi_j + \frac{1}{r} \upgamma^j \Uppsi_4=0\,.
\end{align}
\end{subequations}
The gauge condition $\slash\!\!\!\Uppsi_\mu=0$ is again used the above equations, yielding \cite{Erdmenger:2013thg}
\beq
&&\left(\frac{N^\prime}{2 \sqrt{NA}} \upgamma^5 \!-\! \frac{2 i \omega}{\sqrt{NA}} \upgamma^0 \!+\! \frac{2 i k {}}{r \sqrt{N}} \upgamma^1 \!-\! \frac{2 m\mathbb{I}_{4\times 4}}{\sqrt{N}}\!+\!\frac{2}{r}\upgamma^5\right)\upgamma^5 \Uppsi_4 \nonumber\\&&\!+\! \frac{2 i k {}}{r \sqrt{N}} \Uppsi_1\!+\! \left[\left(\frac{N^\prime}{2 \sqrt{NA}} \!-\!\frac{1}{r}\upgamma^5\right) \upgamma^5 \!-\! \frac{2 i \omega}{\sqrt{NA}} \upgamma^0 \right] \Uppsi_0 \! =\! 0\,.
\label{ccee}
\eeq
Taking into account the  hydrodynamical regime, the gravitino can be expanded to the first order, with respect to $\omega$ and $k$, 
\begin{equation}\label{grav11}
	\Uppsi_\mu = \uppsi_\mu + k\uptau_\mu+ \omega\upvarphi_\mu   \,.
\end{equation}
To the  lowest order terms,  $\omega=0=k$,  in Eq.~\eqref{eom2} and~\eqref{ccee}, it reads 
\begin{equation}
	\uppsi_4^\prime +\left(\frac{3N^\prime}{4 \sqrt{NA}}+ \frac{9}{2r} - \frac{m}{\sqrt{N}}\upgamma^5\right) \uppsi_4 =0\,.
\end{equation}
Analogously, 
\begin{align}
&	\!\!\!\!\!\Uppsi_0^\prime \!+\!\left(\frac{N^\prime}{4 \sqrt{NA}}\!+\! \frac{3}{2r} \!+\! \frac{m}{\sqrt{N}} \upgamma^5\!\right) \Uppsi_0 \!=\! -\frac{N^\prime}{2 \sqrt{NA}}\upgamma^0 \upgamma^5 \uppsi_4\,,\\
&	\uppsi_1^\prime +\left(\frac{N^\prime}{4 \sqrt{NA}} + \frac{3}{2r} + \frac{m}{\sqrt{N}} \upgamma^5 \right) \uppsi_1 = \frac{1}{r} \upgamma^1 \upgamma^5 \uppsi_4\,.
\end{align}
The equation regarding the $\uppsi_4$ component was integrated in Ref. \cite{Erdmenger:2013thg}, after splitting it similarly to~\eqref{esplit}. The other components were obtained and derived in Ref. \cite{Erdmenger:2013thg}, employing the solution for $\uppsi_4$ component. Hereupon the  parameters $a_k, b_k, c_k$ and $d_k$ stand for (integration) constants that arise from the integration of the parameters in Eq. (\ref{esplit}), with respect to the gravitino component $\uppsi_k$ \cite{Erdmenger:2013thg}. 
In the near-horizon regime, one finds 
\begin{align}
	\uppsi_4=&\frac{r_+^{-\frac14}}{2^{3/2} }\begin{pmatrix} \beta a_4 \\ \frac{\beta+1}{2}a_4 \\ \beta c_4 \\ -\frac{\beta+1}{2}c_4 \end{pmatrix}  (r-r_+)^{-\frac34} \,,\label{fl0}\\&
	\overset{\beta \to1}{=}\frac{r_+^{-\frac14}}{2^{3/2}}\begin{pmatrix} a_4 \\ a_4 \\ c_4 \\ -c_4 \end{pmatrix}  (r-r_+)^{-\frac34}.\label{fl1}
\end{align}
On the CFT boundary, ingoing conditions 
impose that the solutions approach $\left(r-r_+\right)^{-\frac{i \omega}{4 \pi T}}$ at the horizon. It implies that $a_4 = b_4$ and $c_4 = - d_4$, in the limit $\beta\to1$. Besides, \begin{align}
	\Uppsi_0&=\frac{r_+^{-1}}{(4r_+)^{3/4} }\begin{pmatrix} \frac{4\beta+1}{5}a_0 \\ \frac{4\beta+1}{5}a_0 \\ \beta c_0 \\ -\beta c_0 \end{pmatrix}  (r-r_+)^{-\frac34}\,,\label{fl2}\\
	&\overset{\beta \to1}{=}\frac{r_+^{-\frac74}}{2^{3/2} }\begin{pmatrix} a_0 \\ a_0 \\ c_0 \\ -c_0 \end{pmatrix}  (r-r_+)^{-\frac34}\,,\label{fl3}\\
	\!\!\!\uppsi_1\!&=\!-\frac{3ir_+^{-\frac{17}{4}}}{2^{3/2}}\!\begin{pmatrix} 3\,{}\beta\, r_+^2 c_4\!-\!2r_+^4c \\ 3{}\beta r_+^2 c_4\!-\!2 r_+^4 c \\ -3{} r_+^2 a_4\!+\!2r_+^4 a \\ 3{} r_+^2 a_4\!-\!2r_+^4 a \end{pmatrix}  (r-r_+)^{-\frac14}\,,\label{fl4}\\
	&\overset{\beta \to1}{=}\!-\frac{3ir_+^{-\frac{17}{4}}}{2^{3/2}}\!\begin{pmatrix} 3r_+^2 c_4\!-\!2r_+^4c \\ 3 r_+^2 c_4\!-\!2 r_+^4 c \\ -3{} r_+^2 a_4\!+\!2r_+^4 a \\ 3{} r_+^2 a_4\!-\!2r_+^4 a \end{pmatrix}  (r-r_+)^{-\frac14}\,.\label{fl5}
\end{align}
Now, we can calculate, for each spinor field, their associated bilinear covariants (\ref{sigma} -- \ref{fierz}) and, subsequently, attribute the spinor field type in the Lounesto's classification (\ref{tipo1} -- \ref{tipo6}). The solutions (\ref{fl0}, \ref{fl2}, \ref{fl4})  compose the gravitino field (\ref{grav11}). Eq. (\ref{fl0}) describes a regular spinor field that flips into a singular, flag-dipole, spinor field, in the $\beta\to1$ limit described by Eq. (\ref{fl1}). On the other hand, the solution (\ref{fl2}) is already a flag-dipole spinor field solution, for any value of $\beta$, according to the protocol in Ref. \cite{Cavalcanti:2014wia}. In particular, the AdS--Schwarzschild limit $\beta\to1$ leads to Eq. (\ref{fl3}). 
In the same way, the spinor field solution (\ref{fl2}) is also a flag-dipole solution, irrespectively of the value of $\beta$.

The AdS boundary behaviour reads  
\begin{align}
\!\!\!(\Uppsi_0^{\upalpha-}\!\!, \Uppsi_0^{\vartheta-}\!\!, \uppsi_1^{\upalpha-}\!\!, \uppsi_1^{\vartheta-})^\intercal
=\frac{2^{4/3}}{3}i  (0,0,c,a)^\intercal r^{-\frac12}\,.
\label{erre}
\end{align}
Defining $	\Uppsi_{\upzeta} =2\upgamma^1 \Uppsi_1 -\upgamma^{3}\Uppsi_{3}- \upgamma^2 \Uppsi_2$, and looking at the equations of motion~\eqref{eom2}, one realizes  that it partly decouples from the other gravitino components:
\beq
	&&\Uppsi_{\upzeta}^\prime+\frac{i\omega}{N}\upgamma^0\upgamma^5\Uppsi_{\upzeta}+\frac{N^\prime}{4 \sqrt{AN}} \Uppsi_{\upzeta} -\frac{i k {}}{r\sqrt{A}}\upgamma^5\left(6\Uppsi_1-\upgamma^1\Uppsi_{\upzeta}\right)\nonumber\\&&\qquad\qquad+\frac{3}{2r} \Uppsi_{\upzeta}-\frac{m}{(NA)^{1/4}}\upgamma^5 \Uppsi_{\upzeta}=0
\eeq
The gauge condition $\slash\!\!\!\Uppsi_\mu=0$ can be then employed to solve the  $\Uppsi_1$ components, 
\begin{equation}
	\Uppsi_1 =\frac{1}{3}\left(\upgamma^1 \Uppsi_{\upzeta} -\upgamma^1\upgamma^0\Uppsi_0 -\upgamma^1 \upgamma^5 \Uppsi_4\right)\,.
	\label{fgh}
\end{equation}
The equations of motion for the $\upvarphi_\mu$ component reads \begin{align}
	&\upvarphi_4^\prime\! +\!\left(\frac{3 N^\prime}{4\sqrt{AN}}+\frac{6}{2r}-\frac{m}{\sqrt{N}} \upgamma^5 \right)\upvarphi_4 \nonumber\\&\qquad\qquad\qquad\qquad=\frac{i}{\sqrt{AN}} \upgamma^0\upgamma^5 \uppsi_4-\frac{2i}{\sqrt{AN}}\Uppsi_0\,,\\
	&\!\!\!\!\!\upvarphi_\zeta^\prime\!+\!\left(\frac{N^\prime}{4 \sqrt{NA}}  \!+\!\frac{3}{2r} \!-\!\frac{m}{\sqrt{N}}\upgamma^5\! \right)\!\upvarphi_\zeta\!=\!-\frac{i}{\sqrt{NA}}\upgamma^0\upgamma^5\uppsi_\zeta\,.
\end{align}
The anti-chiral boundary values are given by 
\begin{align}\label{flagp}
\begin{pmatrix} \upvarphi_0^{\upalpha} \\ \upvarphi_0^{\vartheta} \\ \upvarphi_1^{\upalpha} \\ \upvarphi_1^{\vartheta}\end{pmatrix}
=- 2^{-\frac34} \, r_+^{-3}\begin{pmatrix}  -i a_1 \\   i c_1\\  c_1\\  a_1  \end{pmatrix}  r^{-\frac12}\,.
\end{align}
The spinor field in Eq. (\ref{flagp}) is 
a flagpole, when $a_1$ and $c_1$ are real constants. 

Now, the $\uptau_\mu$ gravitino components are, analogously, given by 
\begin{align}
\begin{pmatrix} \uptau_0^{\upalpha} \\ \uptau_0^{\vartheta} \\ \uptau_1^{\upalpha} \\ \uptau_1^{\vartheta}\end{pmatrix}
={2^{1/4}r_+^{-3}}\begin{pmatrix}
   (1{}- 
   r_+^{-2}/3) \,c_\upzeta\\
  (1{} - r_+^{-2}/3)\, a_\upzeta\\
-3i{}  a\\
3i{}  c
\end{pmatrix}  r^{-\frac12}\,.\label{fp1}
\end{align}
This is clearly a regular spinor field.
If $c=c_\upzeta$, $a=a_\upzeta$,  and $r_+=\sqrt{3}$, then Eq. (\ref{fp1}) is a flagpole singular spinor field, as explicitly 
verified to satisfy the conditions (\ref{tipo5}).
Besides, if either $a_\zeta=0=c_\zeta$ or 
$a=0=c$, then Eq. (\ref{fp1}) satisfies the conditions (\ref{tipo6}), hence regarding a dipole spinor field.

Ref. \cite{Erdmenger:2013thg} shows that the components $\Uppsi_0, \Uppsi_1$ have anti-chiral  boundary values given by 
\begin{align}
(\Uppsi_0^{\upalpha}, \Uppsi_0^{\vartheta}, \Uppsi_1^{\upalpha},  \Uppsi_1^{\vartheta})^\intercal
=\mathfrak{B} (a_4,a,c_4,c)^\intercal r^{-\frac12}\,,
\end{align}
for a matrix $\mathfrak{B}$. Solutions can be evaluated at the poles of the boundary values, being therefore interpreted as  phonino modes \cite{Erdmenger:2013thg}. Computing the  $\det\mathfrak{B}$ and substituting the dispersion relation, 
$
	\omega = v_{\rm s} k - i {\rm D}_{\rm s} k^2
$, and solving for $v_{\rm s}$ and ${\rm D}_{\rm s}$,  yields $
	v_{\rm s}=\frac{1}{3}$ and $2 \pi T {\rm D}_{\rm s} = \frac{4\sqrt{2}}{9},$ 
being equal to Eq. (\ref{sgen}), obtained in another context.

\section{concluding remarks}

The occurrence of flag-dipole fermions 
in physics is very rare,  comprising  
features that approach regular spinor fields, although being singular ones, in  the Lounesto's classification (\ref{tipo1} -- \ref{tipo6}) according the bilinear covariants. Besides the two previous 
examples in the literature, here flag-dipole solutions corresponding to the spinor part of the gravitino field  (\ref{grav11}) were obtained, together with flagpole spinor fields. They were derived 
as solutions of the bulk action for the gravitino field,  in the 
 background of black brane solutions that generalize the AdS--Schwarzschild one.
 Besides, the generalized black brane
 has a free parameter driving the singular spinor field solutions, which can flip between regular and singular spinor fields.
 The relation between hydrodynamic transport coefficients and the universal absorption cross sections in the corresponding gravity dual was also studied in the fermionic sectors of the fluid/gravity correspondence. The Kubo formul\ae\, was employed to derive the transport coefficients for the gravitino and its dual, the phonino. In the limit where the generalized black brane parameter tends to the unit, these results are in full compliance with the ones in Ref. \cite{Erdmenger:2013thg} for the AdS--Schwarzschild black brane. The supersound diffusion constants were also discussed through the solutions of the equations of motion for the gravitino. 

\acknowledgments
PM thanks to CAPES and UFABC.  
RdR~is grateful to the National Council for Scientific and Technological Development -- CNPq (Grant No. 303293/2015-2),
and to FAPESP (Grant No. 2017/18897-8), for partial financial support. The authors thank to A. J. Ferreira--Martins for fruitful discussions.

\end{document}